\documentclass[10pt, conference, letterpaper]{IEEEtran}

\usepackage{amssymb,amsthm,amsbsy}
\usepackage{amsfonts,amsmath}
\usepackage{latexsym}
\usepackage[mathscr]{euscript}
\usepackage{graphics}
\usepackage{graphicx}
\usepackage{cite}
\usepackage{cases}

\newcommand{\beq}{\begin{equation}}
\newcommand{\eeq}{\end{equation}}

\newcommand{\beqq}{\begin{equation*}}
\newcommand{\eeqq}{\end{equation*}}
\newcommand{\ei}{\end{itemize}}
\newcommand{\bi}{\begin{itemize}}

\newtheorem{prop}{Proposition}

\newtheorem{algorithm}{Algorithm}

\newtheorem{corol}{Corollary}

\theoremstyle{remark}

\newtheorem{remark}{Remark}

\newcommand{\argmax}[1]{\arg{\hbox{$\underset{#1}{\max}\,$}}}

\newcommand{\ls}[1]
   {\dimen0=\fontdimen6\the\font \lineskip=#1\dimen0
\advance\lineskip.5\fontdimen5\the\font \advance\lineskip-\dimen0
\lineskiplimit=.9\lineskip \baselineskip=\lineskip
\advance\baselineskip\dimen0 \normallineskip\lineskip
\normallineskiplimit\lineskiplimit \normalbaselineskip\baselineskip
\ignorespaces }
\ls{1}

\begin{document}
\title{Toward Fully Coordinated Multi-level \\Multi-carrier Energy Efficient Networks}

\author{Piotr~Wiecek\authorrefmark{1},~Majed~Haddad\authorrefmark{2},~Oussama~Habachi
\authorrefmark{3} and~Yezekael~Hayel\authorrefmark{3}\\
\authorrefmark{1}Institute of Mathematics and Computer Science, Wroclaw University of Technology, Poland\\
\authorrefmark{2}INRIA Sophia-Antipolis, Sophia-Antipolis, France\\
\authorrefmark{3}CERI/LIA, University of Avignon, Avignon, France
}

\maketitle

\begin{abstract}

Enabling coordination between products from different vendors is a key characteristic of the design philosophy behind future wireless communication networks. As an example, different devices may have different implementations, leading to different user experiences.
A similar story emerges when devices running different physical and link layer protocols share frequencies in the same spectrum in order to maximize the system-wide spectral efficiency. In such situations, coordinating multiple interfering devices presents a
significant challenge not only from an interworking perspective (as a result of reduced infrastructure), but also from an implementation point
of view. The following question may then naturally arise: How to accommodate integrating such heterogeneous wireless devices seamlessly?
One approach is to coordinate the spectrum in a centralized manner. However, the desired autonomous feature of future wireless systems makes the use of a central authority for spectrum management less appealing. Alternately, intelligent spectrum coordination have spurred great interest and excitement in the recent years.
This paper presents a multi-level (hierarchical) power control game where users
jointly choose their channel control and power control selfishly in order to maximize their individual energy efficiency. By hierarchical, we mean that some users' decision priority is higher/lower than the others. We propose two simple and nearly-optimal algorithms that ensure complete spectrum coordination among users. Interestingly, it turns out that the complexity of the two proposed algorithms is, in the worst case, \textbf{\emph{quadratic}} in the number of users, whereas the complexity of the optimal solution (obtained through exhaustive search) is $\pmb{N!}$.
These results offer hope that such simple and accurate power control algorithms can be designed around competition, as hierarchical behavior does not only improve the mechanism's performance but also leads to simpler distributed power control algorithms.

\end{abstract}

\vspace{0.1cm}
\begin{IEEEkeywords}
Spectrum coordination, multi-carrier networks, multi-level system, energy efficiency, power control game.
\end{IEEEkeywords}
\vspace{-0.1cm}

\section{Introduction}\label{sec:intro}

Operators are facing exponential traffic increase in their newly
deployed networks, which is becoming a challenge in terms of
investments and energy consumption \cite{Commag11Green,EE12GT}. Since $2010$, under the GreenTouch consortium initiative \cite{GreenTouch}, cellular network actors have identified different solutions and scenarios of future networks that could
sustain the expected traffic huge demand while maintaining a
good quality of service (QoS) and reducing environmental
impact. As an example, the ambitious mission of the GreenTouch
consortium initiative is to deliver the architecture, specifications and
roadmap in order to increase total network energy efficiency by a factor
of $1000$ compared to $2010$ levels \cite{GreenTouch}. Whether these solutions are based on optimizing the
macro layer, using new frequency bands or introducing small
cells, they all require a careful study in terms of energy
consumption impact and operational costs.

In
future networks such as heterogeneous networks, interference management is a crucial issue since the interference due to spectrum sharing can
significantly degrade the overall performance \cite{survey3GPPHetNet11}. In the existing
works, various resource allocation methods are proposed to
either improve energy efficiency or alleviate interference.
However, very little research has addressed their joint interaction. Furthermore, most existing works are throughput-based-utility
problems which lead to a water-filling power control \cite{tse-book} where only a certain number of carriers are exploited depending on the channel gains. In particular, when the signal-to-noise plus interference ratio (SINR) is low (resp. high), only one (resp. every) carrier is exploited. Energy efficiency, on the other hand, is always maximized by using a single carrier \cite{meshkati-jsac-2006}. This means that an energy efficient transmitter (\emph{e.g.}, a primary user) always leaves $K-1$ bands completely vacant for other users (\emph{e.g.}, secondary users) with $K$ being the total number of carriers.
This makes the coordination feature across the multiple interfering devices particularly appealing, not only from an interworking perspective (as a result of reduced infrastructure), but also for improving both network coverage and capacity without the need to split the available spectrum. As an example, in heterogeneous networks, the coordination between macro cells and small cells can increase capacity in the order of $2$ to $10$ times through spectrum reuse across layers and radio coordination functionality \cite{EricssonWP12}.
The practical performance of spectrum coordination is, however, limited
by a variety of nonidealities, such as insufficient channel knowledge,
high computational complexity, heterogeneous user conditions, transceiver impairments, and the constrained level
of coordination between users.\\

One major motivation of this paper is to study fully coordinated energy efficient models which can
be easily implemented in practice. A key technical mechanism of multi-user resource allocation is that of power control, which serves as a mean for both battery saving and interference management.
We shall address a multi-level (hierarchical) power control game where transmitters choose their control policy selfishly in order to maximize
their individual energy efficiency. We shall further model the system as a decentralized multiple access scheme in the sense that the receiver does not
dictate to the users their transmit power levels. Hence, each user
can choose freely his power control policy in order to selfishly
maximize his individual performance criterion, called utility
(or payoff) in the context of game theoretic studies.

\vspace{0.1cm}

\section*{Related Work}

There have been many works on hierarchical (Stackelberg) games, even in the context
of cognitive radio \cite{bloem-gamecomm-2007}, but they do not
consider energy efficiency for the individual utility as defined in
\cite{goodman-pcomm-2000, meshkati-jsac-2006, meshkati-spmag-2007}.
They rather consider transmission rate-type utilities (see
\emph{e.g.,} \cite{LeeSC05,ElGamal2008} for water-filling games, \cite{basar-jota-2002} for pricing models, \cite{Hierarchical-Game-CR-Dusit-JSAC12} for coalition formation games). Hierarchical games are in fact a mechanism for wireless
networks in which some wireless nodes (leaders) have the priority to access the medium, whereas other users (followers) have a low priority to transmit.
In the literature, energy efficient power
control game has been first proposed by Goodman \emph{et al.} in
\cite{goodman-pcomm-2000} for flat fading channels and re-used by
\cite{meshkati-jsac-2006} and \cite{Buzzi-11-Game-EE-CDMA} for multi-carrier code-division multiple
access (CDMA) systems, \cite{Zappone-EE-CDMA-relay-11} for relay-assisted DS/CDMA and \cite{Bacci-13-Game-EE-OFDMA} for orthogonal frequency-division multiple access (OFDMA) communication systems. All these works did not consider hierarchy among different actors in the system.

\begin{figure}[t]
\centering
\vspace*{-2cm}
\hspace*{-1cm}
\includegraphics[height = 20cm,width=12cm]{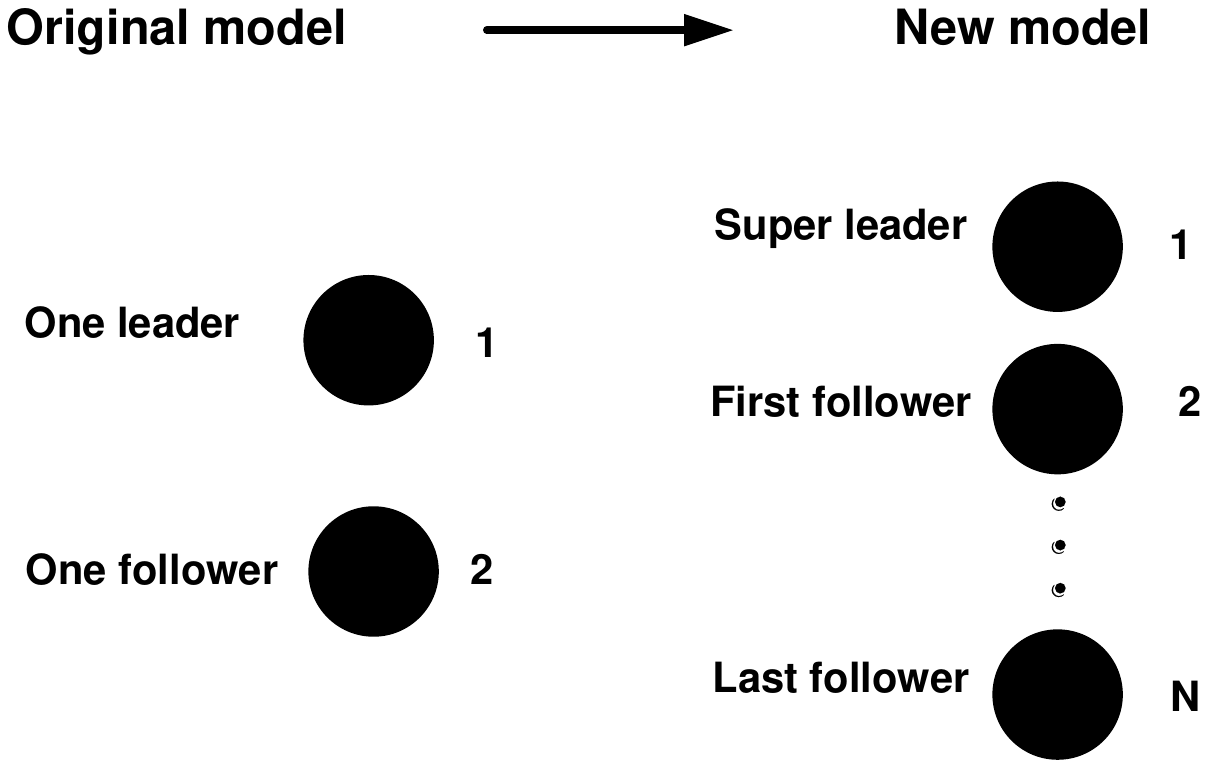}
\vspace{-12.5cm}
\caption{The original model in \cite{Majed-GC-EE} and the new model.}
\label{fig:sys}
\end{figure}

In \cite{Majed-TVT2014}, authors studied a hierarchical game theoretic model for two-user two-carrier energy efficient wireless systems. The power
control problem is modeled as a Stackelberg game where transmitters
choose their control policy selfishly in order to maximize
their individual energy efficiency. Interestingly, it was shown that, under some assumptions, introducing a certain degree of hierarchy in a multi-carrier system results in a
coordination game where the user with the higher priority (\emph{i.e.}, the leader) and the user with the lower priority (\emph{i.e.}, the follower) have incentive to choose their transmitting carriers in such a way that they always transmit on orthogonal channels.

\section*{Contributions}
More specifically, we propose two nearly-optimal though low complex algorithms to achieve spectrum coordination at the equilibrium of the hierarchical game, namely:
\bi
\item The $\delta$-Optimal Complete Spectrum Coordination Algorithm ($\delta$-OCSC Algorithm) where hierarchy is introduced exogenously, \emph{e.g.,} through the order of joining the network,
\item The $\delta$-Modified Complete Spectrum Coordination Algorithm ($\delta$-MCSC Algorithm) where hierarchy is introduced to minimize the distance from the equilibrium based on already known channel gains.
\ei

The main rationale behind the proposed algorithms is that if each entering player maximizes his payoff using all the carriers, he may (and often will) degrade the utility of the players before him. Assuming, that each player chooses only from the carriers left idle by his predecessors, we obtain a number of profits:
\bi
\item no utility degradation by the players joining the game later. In other words, the players profit from being leaders, and thus will not change their decisions after others will join the game,
\item each player knows his final utility when he makes his decision. He does not need to know neither how many players will come after him, nor their channel quality.
\ei

Under this setting, for a large number of players, instead of using a complicated, and not necessarily convergent algorithms like the one of Meshkati \emph{et al.} \cite{meshkati-jsac-2006}, it is reasonable to use the proposed fast and simple algorithms, as in a vast majority of cases it also gives an exact equilibrium solution. These outcomes are meaningful to many practical scenarios. Indeed, it is easy to see that
this hierarchy concept can be directly applied to any network in which users access to the medium in an asynchronous manner by employing a \emph{listen-before-talk} mechanism such as cognitive
radio networks (with the primary user as the leader and secondary users as followers). This is also a natural setting for (multi-tier) heterogeneous wireless networks due to the absence of coordination among the small cells, and between small cells and macro cells \cite{Femto08Survey,Majed_wiopt14}.\\

In the light of the above, the paper is structured as follows. The general system model is
provided in Sec. \ref{sec:model}. Sec. \ref{sec:net-energy-eff} defines the energy efficiency
framework and Sec. \ref{sec:game-model} presents the game theoretic formulation. Then, in Sec. \ref{sec:SE}, we present the two proposed algorithms to achieve complete spectrum coordination at the equilibrium.
Sec.~\ref{sec:simul} provide numerical results to illustrate and validate the
theoretical results derived in the previous sections.
Additional comments and conclusions are provided in Sec.
\ref{sec:conc}.

\section{System Model and Problem Statement}
\label{sec:model}

\subsection{System Model and Notations}
\label{sec:model}

Consider a
decentralized multiple access channel with an arbitrary number of
users $N$ enjoying a slotted bandwidth of $K$ carriers. We assume that the system has $N$ levels of hierarchy, \emph{i.e.}, one user per level as follows: the super
leader (player with the highest priority) is assigned with level
index $1$, and the last follower (player with the lowest priority)
is assigned with level index $N$. Accordingly, interference experienced by user $n$ for $n=\{2,\dots,N\}$ results from the upper levels' transmissions, \emph{i.e.}, users $m \in \{1,\ldots,n-1\}$. It follows then that, for any user $n \in \{1,\ldots,N\}$, the received SINR is expressed as
\beq\label{eq:gamma-mc}
\displaystyle \gamma_{n}^k=\frac{g_{n}^k p_{n}^k}{\sigma^2+ \displaystyle\sum_{\substack{m=1}}^{n-1} g_{m}^k p_{m}^k}:=p_n^k \widehat{h}_n^k
\eeq
We will call $\widehat{h}_n^k$ the \emph{effective channel gain} defined as the ratio between the
SINR and the transmission power for the users over the $k^{th}$ carrier.
$g_n^k$ and $p_{n}^k$ are resp. the fading channel gain and the power control of user $n$ transmitting on carrier $k$, whereas $\sigma^2$ is the variance of Gaussian noise. For realistic reasons and since we try to obtain a complete spectrum coordination, we will further assume that $K\geq N$.

\subsection{Network energy efficiency analysis}
\label{sec:net-energy-eff}

Our system model is based
on the seminal paper \cite{goodman-pcomm-2000} that defines the energy efficiency
framework. To pose the power control problem as a game, we first need to define a utility function suitable for
data applications.
SINR is a critical parameter for the QoS of the signal transmission, as it directly
determines the bit error rate, which is closely related to the
data throughput (average rate of successful packet delivery). In
brief, when SINR is very low, data transmission results in massive
errors and the throughput tends to $0$; when SINR is very
high, data transmission becomes error-free and the throughput
grows asymptotically to a constant. However, achieving a high SINR level requires the
user terminal to transmit at a high power, which in turn results
in low battery life. This phenomenon can
be concisely captured by an increasing, continuous and S-shaped
"efficiency" function $f(\cdot)$, which measures the
packet success rate. Therefore, the throughput of user $n$ over carrier $k$ can
be expressed as
\begin{eqnarray}
\label{eq:def-of-throughput} T_n= R_n\cdot f(\gamma_n^k)
\end{eqnarray}
where $R_n$ stands for the transmission rate of user $n$ and $\gamma_n^k$ stands for the transmission
rate and the SINR of user $n$ over carrier $k$. We also require that to ensure that the utility is equal to $0$ when $p_n=0$. On the other hand, increasing the transmit
power clearly favors the packet success rate and therefore
the throughput. However, as the packet success rate tends to
one, further increasing the power can lead to marginal gains
in terms of throughput regarding the amount of extra power
used. The following utility function allows one to measure
the corresponding trade-off between the transmission benefit
(total throughput over the $K$ carriers) and cost (total power over the $K$ carriers):
\beq\label{eq:util-mc}
u_n(\mathbf{p_1},\ldots,\mathbf{p_{n}})=\frac{\displaystyle R_n \cdot \sum_{k=1}^K  f(\gamma_{n}^k)}{\displaystyle\sum_{k=1}^K p_{n}^k}
\eeq
where $\mathbf{p_{n}}$ is the power control
vector of user $n$ over all carriers, \emph{i.e.}, $\mathbf{p_{n}}=(p_n^1,\ldots,p_n^K)$.
The utility function $u_n$ that has bits per joule as units perfectly captures the trade-off between throughput and battery life and is
particularly suitable for applications where energy efficiency is
crucial.
It should be noted that the throughput in (\ref{eq:def-of-throughput}) could be replaced with
any increasing concave function as long as we make sure that $u_n(0,\ldots,0)=0$. It can be
shown that, for a sigmoidal efficiency function, the utility function in (\ref{eq:util-mc}) is a
quasi-concave function of the user's transmit power \cite{rodriguez-globecom-2003}, and we will use this assumption throughout our paper. This is also
true if the throughput in (\ref{eq:def-of-throughput}) is replaced with an increasing concave
function of $\gamma_n^k$.

\section{The multi-level game theoretic formulation}\label{sec:game-model}
We consider a multi-level game framework in which the leader (referred to hereafter as user $1$) decides first his power control
vector $\mathbf{p_{1}}$ and based on this value, the followers (referred to hereafter as user $n$ for $n=2,\dots,{N}$) will adapt their power
control vectors $\mathbf{p_{n}}$ in a sequential way depending on their level of hierarchy according to the following optimization problem
\beq\label{eq:opt_prob}
\overline{p}_n^k(\mathbf{p_1},\ldots,\mathbf{p_{n}})= \argmax{\mathbf{p_n}} u_n(\mathbf{p_1},\ldots,\mathbf{p_{n}})
\eeq

To characterize the solution of the above problem in a multi-level system, we shall first look at the
last follower (user $N$) who makes the first move to decide his transmit power. An equilibrium can thus be determined using a multi-level approach, where, given the action of higher users in the hierarchy, we compute the best-response function of the follower (the function
$\overline{p}_n(\cdot)$ for $n=2,\dots,N$) and find the actions of the followers which maximize their utilities. We characterize this best-response function by using a result from
\cite{meshkati-jsac-2006}.

This result comes directly from Proposition $1$ of \cite{meshkati-jsac-2006} which we adapt for the multi-level hierarchical game.

\begin{prop}  \label{prop:follower-power}
Given the power control vector $\mathbf{p_1},\ldots, \mathbf{p_{n-1}}$ of the first $n-1$ players in the hierarchy, the best-response of the player $n$ (for $n \in \{2,\ldots,N\}$) is
given by
\small
\begin{equation}\label{eq:follower}
\overline{p}_n^k= \left\{\begin{array}{lr}\displaystyle
\frac{\gamma^{*}(\sigma^2+\sum_{m=1}^{n-1}g_{m}^k p_{m}^k)}{g_{n}^k}, \,\,\, \mbox{for} \, k = L_n(\mathbf{p_1},\ldots,\mathbf{p_{n-1}}),\\
0, \qquad\qquad\qquad\qquad \qquad \mbox{for all}\,\,  k \neq L_n(\mathbf{p_1},\ldots,\mathbf{p_{n-1}})
\end{array}
\right.
\end{equation}
\normalsize
with $L_n(\mathbf{p_1},\ldots,\mathbf{p_{n-1}})=\argmax{k} \widehat{h}_n^k(\mathbf{p_1},\ldots,\mathbf{p_{n-1}})$ and $\gamma^*$ is the unique
(positive) solution of the first order equation
\begin{equation}\label{eq:gamma*}
x\,f^{\prime}(x)=f(x).
\end{equation}
\end{prop}

Equation (\ref{eq:gamma*}) has a unique solution if the efficiency function $f(\cdot)$ is
sigmoidal \cite{rodriguez-globecom-2003}.
The last proposition says that the best-response of the follower is to use only one carrier, the
one such that the effective channel gain is the best.\\

One could wonder why not let each player choose its power allocation strategy according to Proposition \ref{prop:follower-power}? The response is that, if a user had all the data, in order to profit from being the leader in that case, he would have to make very complex computations, whose complexity would grow exponentially with the number of remaining players. To overcome these computational challenges among others, we present, in the next section, two simple nearly-optimal algorithms to obtain complete spectrum coordination among users.

\section{Proposed Spectrum Coordination Algorithms}\label{sec:SE}

In this section, we will show that using a simple procedure combined with a well chosen hierarchy among the players, we may obtain both a complete spectrum coordination and payoffs not far from the optimum for each of the players.
This can be achieved in two different scenarios. In a way, considering a centralized mode where the proposed
system would require information from a third party (\emph{i.e.}, central database maintained by regulator
or another authorized entity) to schedule users coming. In another way, an extra signaling channel is dedicated to perform the collision detection so that users
will not transmit at the same moment. Such an assumption could be justified by the fact that in an asynchronous context, the
probability that two users decide to transmit at the same moment is
negligible as the number of users is limited.

We start by defining a \emph{naive} mechanism\footnote{As we will see later in the paper, Algorithm 1 is just a starting point for Algorithms 2 -- 4.} for determining the power control for all the players. Let $\pi$ be an arbitrary ordering (permutation) of the players.
\begin{algorithm}\label{alg:CSC}
$\pi$-Complete Spectrum Coordination Algorithm ($\pi$-CSC Algorithm)
\\
For $n=1,\ldots,N$ do the following steps:
\begin{enumerate}
\item For player $\pi(n)$, choose the carrier with the highest $g_{\pi(n)}^k$ that is not already chosen by some other player. Denote it by $\kappa(n)$.
\item Choose the following power control for player $\pi(n)$:
$$\overline{p}_{\pi(n)}^k=\left\{ \begin{array}{ll}\frac{\gamma^*\sigma^2}{g_{\pi(n)}^k},&\mbox{when }k=\kappa(n)
\\0,&\mbox{otherwise}\end{array}\right.$$
\end{enumerate}
\end{algorithm}
The proposition below gives some important features of this algorithm. To present it, we need to introduce some additional notations.
For each player $n$ and every carrier $k$ define $\rho_n^k=\frac{g_n^k}{g_n^{B_n}}$ where $B_n$ is the best channel for player $n$. This fraction can be interpreted as loss of quality ratio of player $n$ from choosing carrier $k$ instead of his best carrier (note that if we assume that each of the players uses different carrier, as we do here, their optimal utilities on their chosen carriers will be $\frac{f(\gamma^*)g_n^kR_n}{\gamma^*\sigma^2}$, which is proportional to $g_n^k$). Next for each $n$ and $l=1,\ldots,K$ let $\rho_n(l)$ be the $l$-th biggest value of $\rho_n^k$.
\begin{prop}
\label{prop:CSC}
If there exists an $\alpha$ such that, for $n=1,\ldots,N$, $\rho_{\pi(n)}(n)>\alpha$, then the $\pi$-CSC Algorithm gives each user the utility not smaller than $\alpha$ times the biggest utility that he can obtain\footnote{By the biggest utility a player can obtain we mean utility that he obtains when all the players cooperate to maximize his utility. It bounds from above both his utility at equilibrium and at the social optimum.}.
\end{prop}
The proof of Prop. \ref{prop:CSC} is given in Appendix \ref{app:CSC}.
Two immediate consequences of this proposition are the following:
\begin{corol}
\label{CSC:equilibrium}
If $\alpha$ satisfies the assumptions of Proposition \ref{prop:CSC}, then the power control defined by the $\pi$-CSC Algorithm are a
$\frac{1-\alpha}{\alpha}$-equilibrium in our power control game.
\end{corol}
\begin{corol}
\label{CSC:soc_opt}
If $\alpha$ satisfies the assumptions of Proposition \ref{prop:CSC}, then the optimal solution of our game when they use power control defined by the $\pi$-CSC Algorithm, is not smaller than $\alpha$ times the optimal solution at the social optimum.
\end{corol}
As we can see, even this simple control mechanism possesses some useful characteristics. It can notably be applied in case the hierarchy is introduced exogenously, such as in asynchronous systems where the probability that two users decide to transmit at the same moment is
negligible, provided that the number of users is limited.

\subsection{The $\delta$-Optimal Complete Spectrum Coordination Algorithm ($\delta$-OCSC Algorithm)}

If we can impose hierarchy between the players in order to improve the performance of the mechanism, we may use the algorithm defined below.
Its main goal will be to find the hierarchy of the players $\pi$ minimizing $\alpha$ (and in consequence giving the solution closest both to the equilibrium and the social optimum in our game). A small value $\delta>0$ is the parameter of the algorithm.

\begin{algorithm}
\label{alg:alpha_pi_choice}
Start with $\underline{\alpha}=0$, $\overline{\alpha}=1$, $\alpha^*=0$ and $\pi^*=[1 \ldots N]$.\\
Repeat the following steps until the procedure is interrupted in point 1):

\begin{enumerate}
\item If $\overline{\alpha}-\underline{\alpha}<\delta$ or $\underline{\alpha}>\frac{1}{1+\gamma^*}$ stop. Otherwise take $\alpha^*=\frac{\overline{\alpha}+\underline{\alpha}}{2}$ and $\pi=\boldsymbol{0}_{1\times K}$.
\item For $n=1,\ldots,N$ do the following steps:\\
Find the smallest $\rho_n(l^*)$ such that $\rho_n(l^*)\geq\alpha^*$. If $\pi(l^*)=0$ then put $\pi(l^*)=n$. Otherwise, find the biggest $l<l^*$ such that $\pi(l)=0$, and put $\pi(l)=n$. If $\pi(l)\neq 0$ for every positive $l<l^*$ put $\overline{\alpha}=\alpha^*$ and return to point 1).
\item Put $\pi^*=\pi$ with all the zero elements removed, $\underline{\alpha}=\alpha^*$ and return to point 1).\\
\end{enumerate}
\end{algorithm}

Now, we can state Algorithm 3.\\

\begin{algorithm}\label{alg:OCSC}
 $\delta$-Optimal Complete Spectrum Coordination Algorithm ($\delta$-OCSC Algorithm)
\begin{enumerate}
\item Compute the values of $\rho_n(l)$.
\item Perform Algorithm \ref{alg:alpha_pi_choice} for $\delta$ and $\rho_n(l)$.
\item If the output $\pi^*=\boldsymbol{0}$, put $\alpha^*=0$ and $\pi^*=[1 \ldots N]$. Then for the output $\pi^*$ perform $\pi^*$-CSC Algorithm.
\end{enumerate}
\end{algorithm}
Given the interpretation of $\rho_n^k$, the parameter $\alpha$ appearing in algorithms above can be interpreted as the maximal loss of utility for any player from not being the leader, that is the worst-case\footnote{Worst-case here means that such a big disutility will only be perceived by the players if their private ordering (from best to worst) of the carriers is similar, e.g. when channel gains for different users are strongly correlated.} ratio of utility of any of the players who do not choose their carriers first to their utility if they were the first ones to choose. The objective of Algorithm \ref{alg:alpha_pi_choice} is to find the ordering of the players which minimizes this disutility.
It is done by putting on $i$-th coordinate of ordering $\pi$ a player (his index), who has at least $i$ good carriers to choose from (by which we mean $i$ carriers with utility better than $\alpha$ times his best possible utility if he was a leader). $\alpha^*$ obtained by Algorithm \ref{alg:alpha_pi_choice} is the minimum value for which such an ordering is possible. As a consequence, the outcome of the $\delta$-OCSC Algorithm will have several useful properties.
\begin{prop}
\label{prop:OCSC}
The choice of power control done by the $\delta$-OCSC Algorithm satisfy the following:
\begin{enumerate}
\item It gives each of the players the utility not smaller than $\alpha^*$ times the biggest utility that he can obtain.
\item It is a $\frac{1-\alpha^*}{\alpha^*}$-equilibrium in the game.
\item The optimal solution in our game when they use power control defined by it is not smaller than $\alpha^*$ times the optimal solution at the social optimum.
\item If $\alpha^*>\frac{1}{1+\gamma^*}$ then the power control given by the outcome of the algorithm is an equilibrium of the game.
\end{enumerate}
Moreover, for $\delta$ small enough there exist the values $\tilde{g}_n^k$ giving the same values of $\rho_n(l)$ as $g_n^k$ such that the power control obtained by the $\delta$-OCSC Algorithm for the game with channel gains equal to $\tilde{g}_n^k$ are the best $\epsilon$-equilibrium with complete spectrum coordination.
\end{prop}
The proof of Prop. \ref{prop:OCSC} is given in Appendix \ref{app:OCSC}.
We need to know that Proposition \ref{prop:OCSC} does not give exact information about the quality of the algorithm proposed, just an upper bound which may be far from exact.
Especially, when $N \ll K$, the bounds on the quality of the solution obtained can be far from reality. In fact, the solution obtained then may either be an equilibrium (or very close to it, even if $\alpha^*$ is smaller than $\frac{1}{1+\gamma^*}$) or almost $\frac{1-\alpha^*}{\alpha^*}$ away from equilibrium, even though there exists one in the game. Anyway, as the last part of the proposition states, the $\delta$-OCSC Algorithm is the best algorithm for choosing the power control of the players, in a certain class. This class consists of the algorithms inducing complete spectrum coordination, and such that the ordering of the players is completely done before the power control starts.\\

\begin{remark}
It is worth noting here that, in case $N \ll K$, a simpler algorithm may give comparable expected utilities to players: instead of finding optimal ordering of the players, we may randomize it. This approach has however two important disadvantages. First, even when $N$ is small in comparison to $K$, with some small probability a very bad ordering may be obtained, which is not possible in $\delta$-OCSC Algorithm. Second, in general it may happen (and the probability of such a situation grows as $N$ approaches $K$) that the output of Algorithm \ref{alg:OCSC} will be optimal in any meaningful sense, while randomization of the ordering of the players followed by Algorithm \ref{alg:CSC} will give with probability close to $1$ utility close to $0$ for some players. An example of such situation is the following: let
$$g_n^k=\left\{ \begin{array}{ll} 1-k\varepsilon,&k\leq n,\\(K-k)\varepsilon,&k>n\end{array}\right.$$
for some small value of $\varepsilon>0$. The optimal ordering of the players here is $1,2,\ldots,n$, with power allocations chosen by the $\delta$-OCSC Algorithm giving each of the players not less than $1-n\varepsilon$ times his maximum utility. On the other hand, when $K=N$, any other ordering will give at least one player no more than $K\varepsilon$ times his maximum utility, which can be arbitrarily small for appropriately chosen $\varepsilon$. We will further discuss this issue in Section \ref{sec:simul} by comparing our approach to a random coordination algorithm.
\end{remark}

\subsection{The $\delta$-Modified Complete Spectrum Coordination Algorithm ($\delta$-MCSC Algorithm)}

To go further with the analysis, and with the same goal of obtaining a better energy efficiency at the equilibrium, we will need an algorithm which updates the ordering of the players after each power control is made. Such a modification should be most effective exactly in case $N \ll K$, when most of the desired power control (that is, the ones giving the players the highest utilities possible) will not interfere with each other, as $\delta$-OCSC Algorithm tries to provide the best ordering in case most of the desired choices do interfere. The pseudo-code for the proposed approach is given in the next algorithm.

\begin{algorithm}\label{alg:MCSC}
$\delta$-Modified Complete Spectrum Coordination Algorithm ($\delta$-MCSC Algorithm)
\\
Compute the values of $\rho_n^k$.
Define the sets of indices $S=\{ 1,\ldots, N\}$ and $\tilde{S}=\{ 1,\ldots,K\}$. For $n=1,\ldots,N$ do the following steps:
\begin{enumerate}
\item For the set of players reduced to $S$ and set of carriers reduced to $\tilde{S}$ update $\rho_n(l)$\footnote{But without updating the values of $\rho_n^k$, just reducing the sets of indices $n$ and $k$.} and perform Algorithm \ref{alg:alpha_pi_choice} for $\delta$ and $\rho_n(l)$.
\item For the output $\pi^*(1)$ choose the carrier $k\in \tilde{S}$ with the highest $g_{\pi^*(1)}^k$. Denote it by $\kappa$.
\item Choose the following power control for player $\pi^*(1)$:
$$\overline{p}_{\pi^*(1)}^k=\left\{ \begin{array}{ll}\frac{\gamma^*\sigma^2}{g_{\pi^*(1)}^k}&\mbox{when }k=\kappa\\0&\mbox{otherwise}\end{array}\right.$$
\item Change the sets $S$ and $\tilde{S}$ according to: $S:=S\setminus \{ \pi^*(1)\}$, $\tilde{S}:=\tilde{S}\setminus\{ \kappa\}$.
\end{enumerate}
\end{algorithm}

In fact, hierarchy between decisions players can be introduced here in order to minimize the gap between the output of the proposed algorithms and the equilibrium (based on already known channel gains). We do not present any formal proposition on the quality of the solution obtained, as all the results of the Proposition \ref{prop:OCSC} still hold for this modification of the algorithm, while in general no better boundaries can be obtained, as the differences between the two algorithms depend strongly on the relation between $N$ and $K$.\\
\begin{remark}
Being higher in hierarchy is, unlike \cite{gaoning_infocom11}, profitable for the players. It is straightforward to see that in Algorithm \ref{alg:CSC} a player higher in hierarchy has more options to choose from, so he can never lose on it. On the other hand, in Algorithms \ref{alg:OCSC} and \ref{alg:MCSC} the hierarchy is chosen to get solution of the game as close to the equilibrium as possible. Consequently, if one of these algorithms finds an exact equilibrium (we will see in the next section that, for the vast majority of cases, we obtain an exact equilibrium solution) a player trying to violate the ordering introduced by the algorithm by going up the hierarchy may lose on it, as this may move the game out of the equilibrium, and responses of other players may degrade the deviator's performance.
\end{remark}

\section{Simulation Results}\label{sec:simul}

We consider the widely used energy efficiency function in power control games, that is $f(x)=(1-e^{-x})^M$, where $M=100$ is the block length in bits. For this efficiency function, $\gamma^*\simeq 6.47$ (or $8.1$ dB). Note that we have simulated $10000$ scenarios to remove the random effects from the Rayleigh fading channels $g_n^k$. We have considered that $SNR=\frac{1}{\sigma^2}$ and that the rate of each user $n$ is $R_n=1$ Mbps. Unless otherwise stated, we fix the number of carriers $K$ to the number of users $N$.

\begin{figure}[t]
\centering
\vspace*{-4cm}
\hspace*{-1.8cm}
\includegraphics[height = 12cm,width=12.2cm]{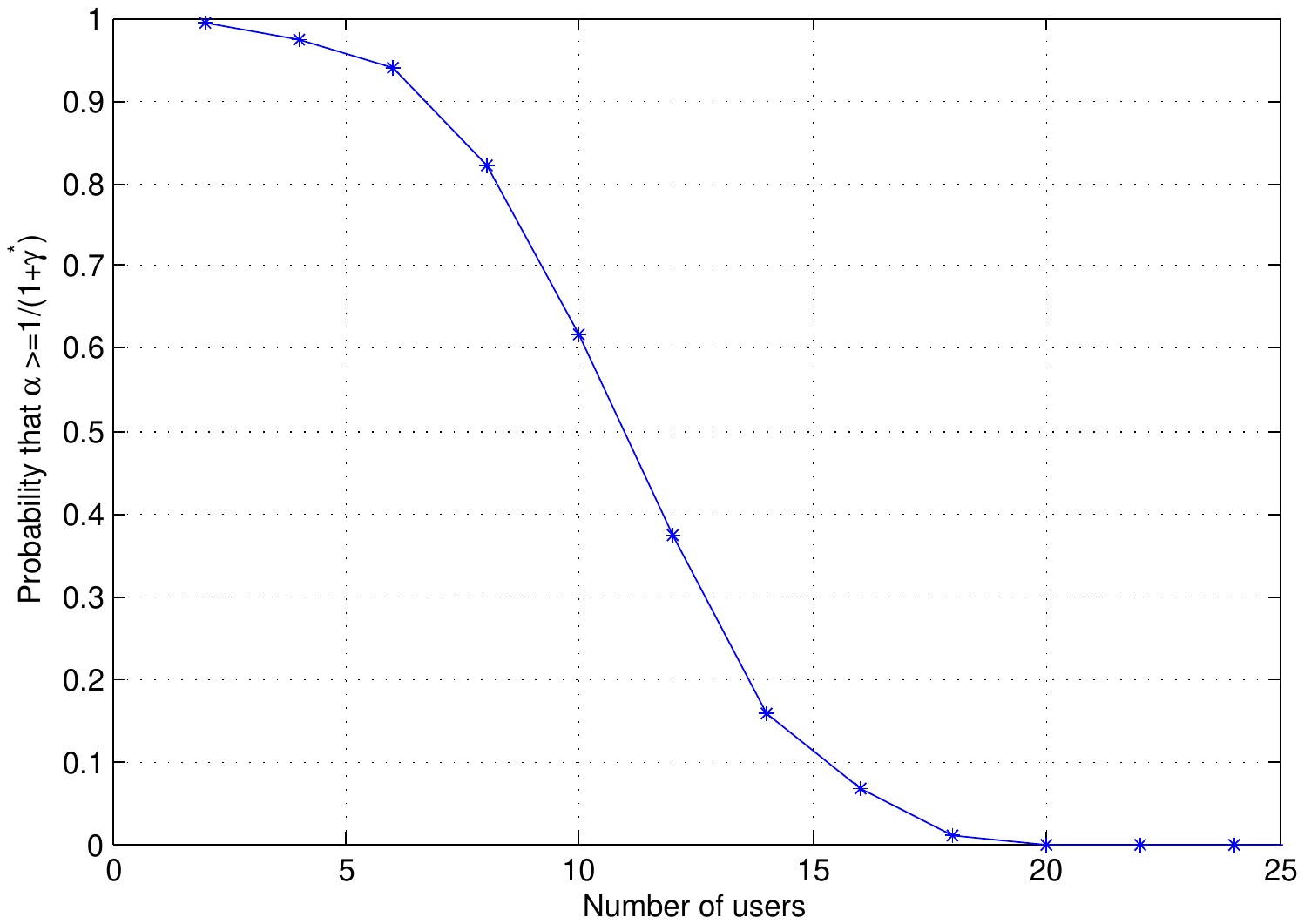}
\vspace{-4cm}
\caption{The probability that $\alpha >=1/(1+\gamma^*)$.}
\label{fig:compt}
\end{figure}

Figure \ref{fig:compt} depicts the probability that $\alpha >=1/(1+\gamma^*)$ which, as part 4) of Proposition \ref{prop:OCSC} claims, guarantees that the power control given by the outcome of the algorithm $\delta$-OCSC
is an exact equilibrium. Interestingly, although it is an \emph{approximate} algorithm, with $98\%$, the $\delta$-OCSC Algorithm finds the value of an \emph{exact} equilibrium, and not an $\epsilon$-equilibrium. This is not unexpected, as by part 4) of Proposition \ref{prop:OCSC}, already the value of $\alpha^*=\frac{1}{1+\gamma^*}$ (which in our case equals to $0.13$) implies that. It is important however to note that this result suggests that, for a big number of players, it is reasonable to use the proposed fast and simple algorithm instead of using a complicated, and not necessarily convergent algorithms like the one of Meshkati \emph{et al.} \cite{meshkati-jsac-2006}, as in a vast majority of cases it gives an exact equilibrium solution.

\begin{figure}[t]
\vspace*{-6cm}
\hspace*{-2cm}
\includegraphics[height = 16cm,width=12.5cm]{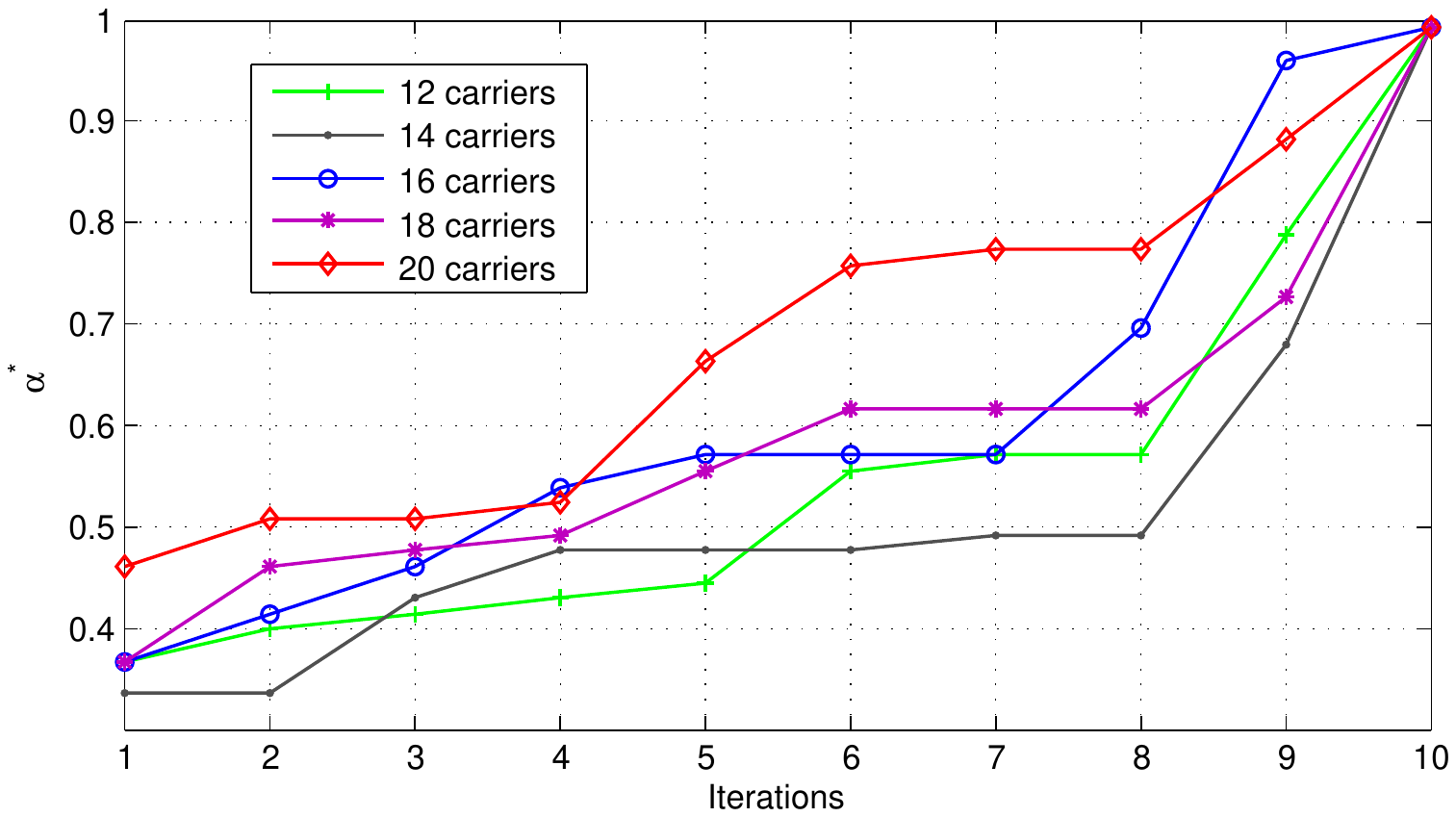}
\vspace{-6cm}
\caption{$\alpha^*$ depending on the number of iterations in Algorithm 4.}
\label{fig:alphas}
\end{figure}

\begin{figure}[t]
\centering
\vspace*{-5cm}
\hspace*{-2cm}
\includegraphics[height = 14.5cm,width=12.8cm]{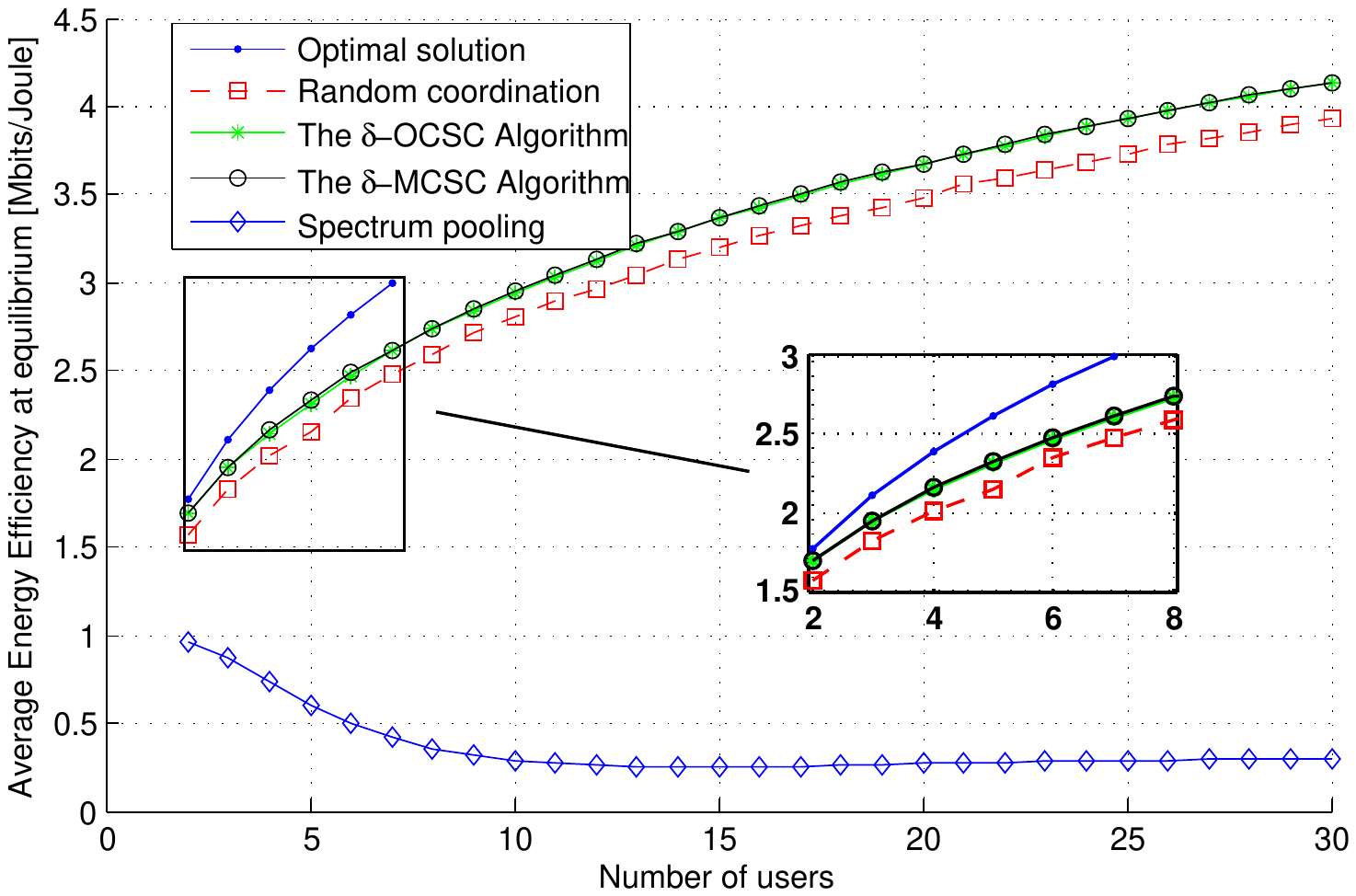}
\vspace{-5cm}
\caption{Average energy efficiency at the equilibrium for increasing number of users for SNR $=10$ dB.}
\label{fig:algoEE}
\end{figure}

We plot in Figure \ref{fig:alphas} the value of $\alpha^*$ for the $\delta$-MCSC Algorithm as a function of iterations for a number of users $N=10$.
Obviously, the first value obtained in Fig. \ref{fig:alphas} is the one corresponding to the $\delta$-OCSC Algorithm. As expected, we observe that the value of $\alpha^*$ increases as the number of iterations increases. Note however, that this increase does not usually imply the increase of energy efficiency as we will see next in Fig.~\ref{fig:algoEE}. But, as we can see in Fig. \ref{fig:alphas}, also in this case, for $K=12$ carriers, we observe a significant increase of $\alpha^*$ from the $6$th iteration on, which can result in the $\delta$-MCSC Algorithm outperforming the $\delta$-OCSC Algorithm.

Let us now illustrate the impact of such parameters on the system performances at the equilibrium. Figure \ref{fig:algoEE} depicts the average energy efficiency when users implement the $\delta$-OCSC Algorithm and the $\delta$-MCSC Algorithm for SNR $=10$ dB, where the average is done over the $N$ users. For our comparison purpose, we exemplify our general analysis by investigating the following algorithms:
\bi
\item \emph{\textbf{the random coordination algorithm}}: which uses a random permutation of users and then applies Algorithm 1,
\item \emph{\textbf{the spectrum pooling algorithm}}: which is throughput-based-utility \cite{Jondral_pooling}, and
\item \emph{\textbf{the optimal solution}}: obtained through exhaustive search.
\ei
We vary the number of players from $2$ (\emph{i.e.}, one leader and one follower) to $30$ (\emph{i.e.}, one leader and $29$ followers).
We first observe that, for both $\delta$-OCSC and $\delta$-MCSC algorithms, the energy efficiency increases with the number of users, which is somewhat intuitive, as a bigger number of players means that more of them will be able to use their best carriers due to channel diversity gain. Another important result is that the two proposed algorithms perform very closely to the optimal solution, which offer hope that our algorithms perform reasonably well trying to make up for the decreasing carrier diversity by using the player diversity instead. Such a result is even more important, as the complexity of the two proposed algorithms is in the worst case \textbf{\emph{quadratic}} in the number of users $N$, whereas the complexity of the exhaustive search is $\pmb{N!}$. Moreover, it is clearly shown that the $\delta$-OCSC and $\delta$-MCSC algorithms outperform the random coordination algorithm as already stated by Remark 1. Finally, we remark that the energy efficiency of the spectrum pooling algorithm decreases as the number of users increases which suggests that throughput-based-utility schemes are even less energy efficient as the number of users increases, because in such case, users have to transmit at higher power as the opportunity to transmit is lower.

\begin{figure}[t]
\centering
\vspace*{-5cm}
\hspace*{-1.8cm}
\includegraphics[height = 14.5cm,width=12.5cm]{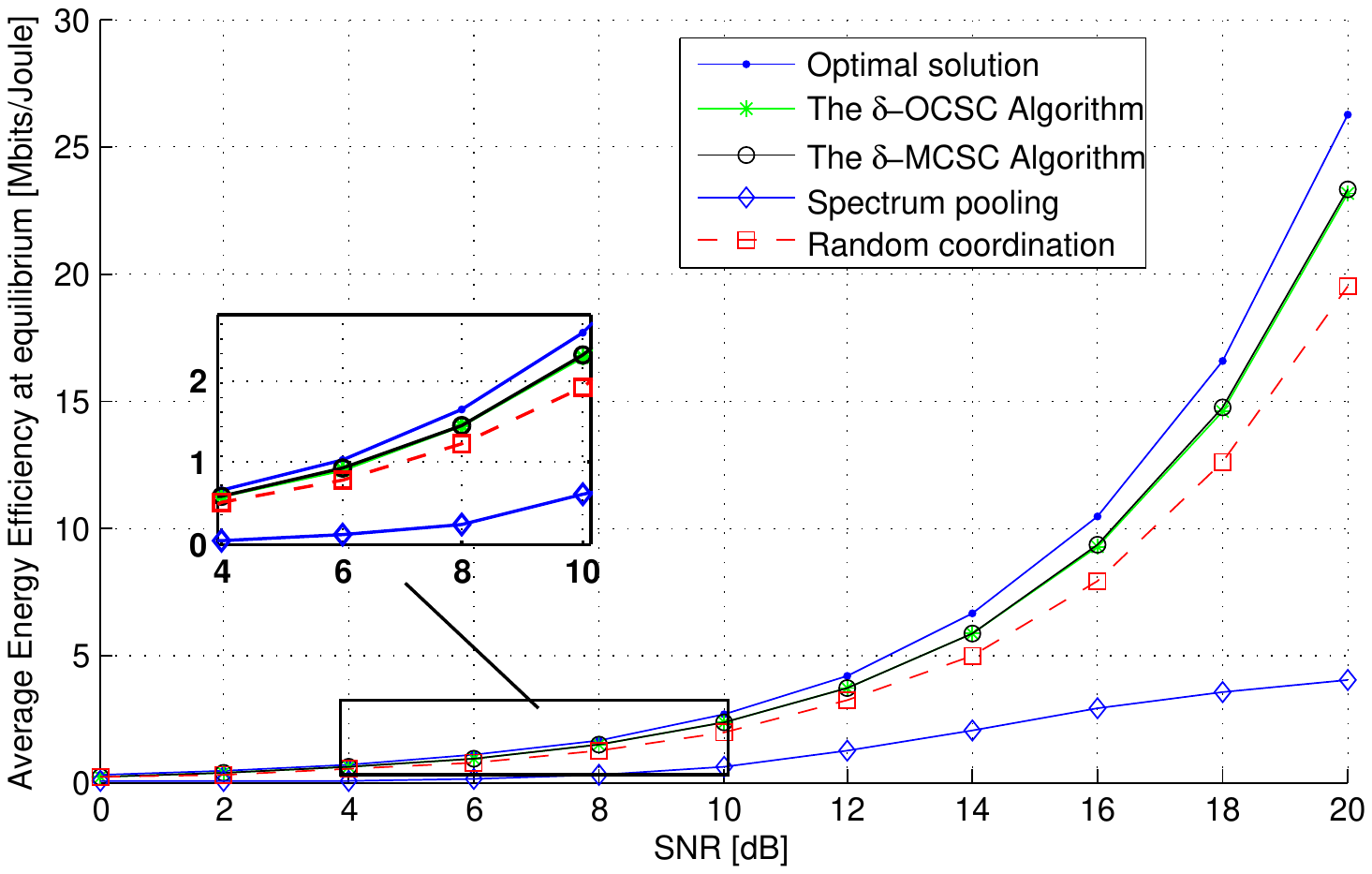}
\vspace{-5cm}
\caption{Average energy efficiency at the equilibrium as a function of the SNR for $N=5$ users.}
\label{fig:EE_SNR}
\end{figure}

In Figure \ref{fig:EE_SNR}, we plot the average energy efficiency at the equilibrium depending on the SNR for $N=5$ users. Again, we observe that the $\delta$-OCSC and $\delta$-MCSC algorithms outperform the random coordination algorithm, and the gap between the two proposed algorithms and the optimal solution is very small even when the SNR increases. Here, the $\delta$-MCSC algorithm performs slightly better than the $\delta$-MCSC algorithm. 
As the SNR decreases, all configurations tend towards having the same (zero) energy efficiency. This can be justified by the fact that, at low SNR regime, whatever the power control strategy each user chooses, the signal is overwhelmed by the noise.

\begin{figure}[t]
\centering
\vspace*{-4.5cm}
\hspace*{-1.8cm}
\includegraphics[height = 14cm,width=12.5cm]{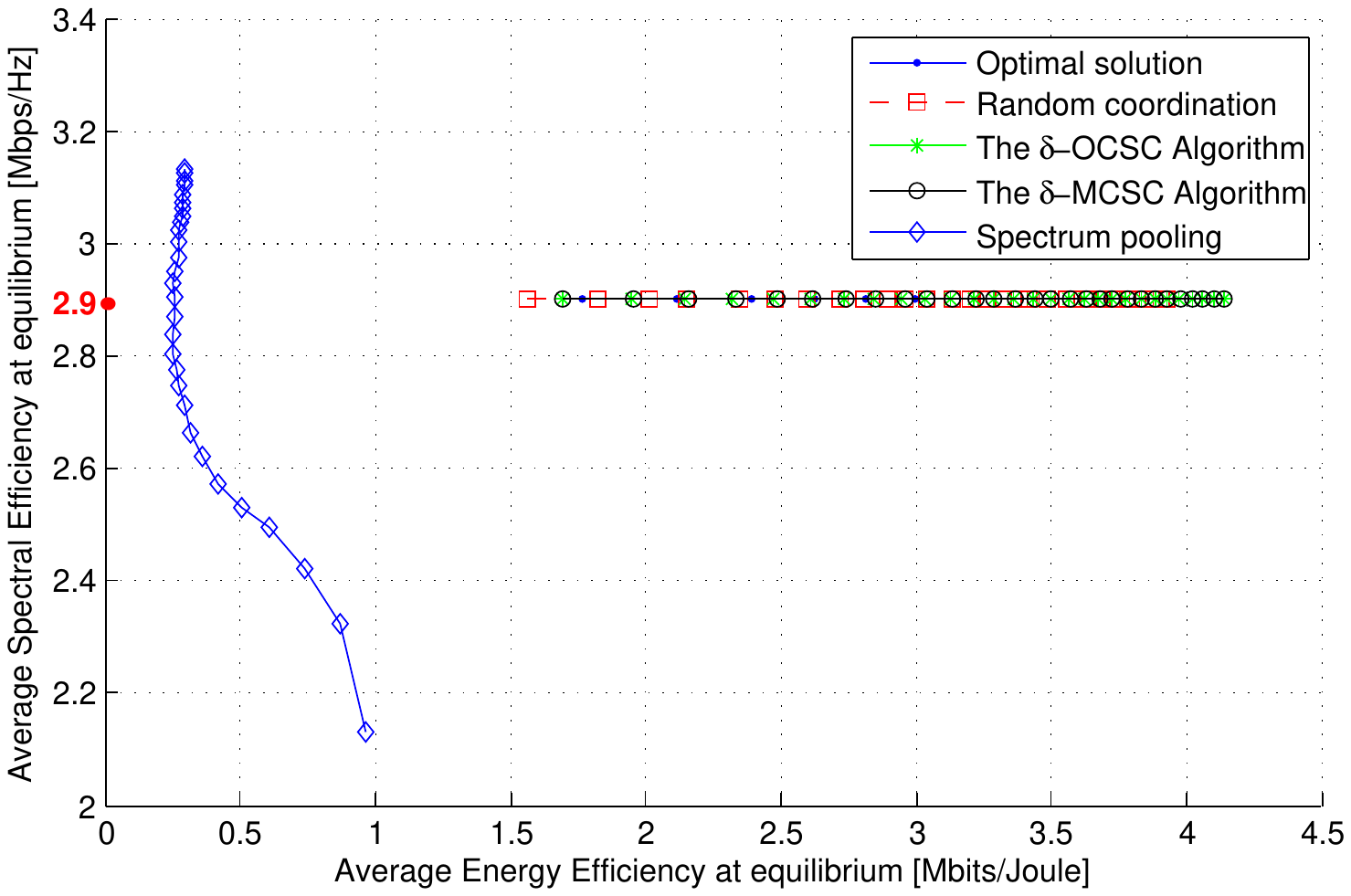}
\vspace*{-5cm}
\caption{Spectral efficiency -- Energy efficiency trade-off.}
\label{fig:se_ee}
\end{figure}

In order to figure out the balance between the achievable rate and energy consumption of the system, we plot in Fig.~\ref{fig:se_ee} the spectral efficiency as a function of the energy efficiency. Particularly noteworthy is the fact that the $\delta$-OCSC and $\delta$-MCSC algorithms optimize the energy efficiency while still maximizing the spectral efficiency, whereas the spectrum pooling scheme performs slightly better in terms of spectral efficiency, but is clearly energy inefficient with the respect to the other schemes. Moreover, the spectral efficiency at the equilibrium of the two proposed algorithms can be computed exactly: It is equal to $\log_2(1+\gamma^*) \simeq 2.9$, with $\gamma^* = 6.47$, which corresponds exactly to the spectral efficiency in Figure \ref{fig:se_ee}. Notice that this contrasts with most related works so far in which the optimal
energy efficiency performance often leads to low spectral efficiency performance and vice versa \cite{tradeoff_green_commag11,MukherjeeM14,DengRCZZL13,wcnc/AminMEH12,JoungESTJSAC2014,TsilimantosGJ13}.
This key features have a great impact on the network performance, and may further guide the design of power control strategies in a multi-carrier energy efficient systems.

\section{Conclusion}\label{sec:conc}

The potential spectral efficiency gain along with the reduced computational complexity make spectrum coordination across multiple interfering users much more challenging than traditional setups built on strict interference
avoidance. In this work, we have considered a hierarchical multiple access multi-carrier system in which users wish to selfishly maximize their energy efficiency by means of adequately choosing their power control policy over the carriers. A main challenge of applying the idea of hierarchical games in the energy efficient multi-carrier context is that, contrary to throughput-based-utility problems where only a certain number of carriers are exploited depending on the channel gains, energy efficient systems always leave $K-1$ bands completely vacant for other users to be exploited. Following the above trend, we have explored a multi-level (hierarchical) energy efficient game framework where all players attempt to coordinate their transmission (in terms of spectrum) and analyze how users behave in the presence of hierarchy. The proposed approach goes toward the vision of a fully coordinated multi-carrier energy efficient wireless network, whereby transmit powers are coordinated across the users.
We have presented two intuitive, easy to implement and
nearly-optimal algorithms which, in $98\%$ of cases, converge to an exact equilibrium. 
Interestingly, our simulation results have shown that the two proposed algorithms achieve a flexible and desirable trade-off between energy efficiency and throughput maximization by exploiting the advantageous intrinsic features of the spectrum coordination paradigm.

%

\bibliographystyle{IEEEtran}
\bibliography{C:/Users/mhaddad/Dropbox/mybib}

\begin{thebibliography}{10}
\providecommand{\url}[1]{#1}
\csname url@samestyle\endcsname
\providecommand{\newblock}{\relax}
\providecommand{\bibinfo}[2]{#2}
\providecommand{\BIBentrySTDinterwordspacing}{\spaceskip=0pt\relax}
\providecommand{\BIBentryALTinterwordstretchfactor}{4}
\providecommand{\BIBentryALTinterwordspacing}{\spaceskip=\fontdimen2\font plus
\BIBentryALTinterwordstretchfactor\fontdimen3\font minus
  \fontdimen4\font\relax}
\providecommand{\BIBforeignlanguage}[2]{{%
\expandafter\ifx\csname l@#1\endcsname\relax
\typeout{** WARNING: IEEEtran.bst: No hyphenation pattern has been}%
\typeout{** loaded for the language `#1'. Using the pattern for}%
\typeout{** the default language instead.}%
\else
\language=\csname l@#1\endcsname
\fi
#2}}
\providecommand{\BIBdecl}{\relax}
\BIBdecl

\bibitem{Commag11Green}
C.~Han, T.~Harrold, S.~Armour, I.~Krikidis, S.~Videv, P.~M. Grant, H.~Haas,
  J.~Thompson, I.~Ku, C.-X. Wang, T.~A. Le, M.~Nakhai, J.~Zhang, and L.~Hanzo,
  ``Green radio: radio techniques to enable energy-efficient wireless
  networks,'' \emph{IEEE Communications Magazine}, vol.~49, no.~6, pp. 46--54,
  2011.

\bibitem{EE12GT}
S.~Zarifzadeh, N.~Yazdani, and A.~Nayyeri, ``Energy-efficient topology control
  in wireless ad hoc networks with selfish nodes,'' \emph{Comput. Netw.},
  vol.~56, no.~2, pp. 902--914, Feb. 2012.

\bibitem{GreenTouch}
\BIBentryALTinterwordspacing
{GreenTouch initiative}. [Online]. Available: \url{www.greentouch.org}
\BIBentrySTDinterwordspacing

\bibitem{survey3GPPHetNet11}
A.~Damnjanovic, J.~Montojo, Y.~Wei, T.~Ji, T.~Luo, M.~Vajapeyam, T.~Yoo,
  O.~Song, and D.~Malladi, ``A survey on 3gpp heterogeneous networks,''
  \emph{Wireless Communications, IEEE}, vol.~18, no.~3, pp. 10--21, 2011.

\bibitem{tse-book}
D.~Tse and P.~Viswanath, \emph{Fundamentals of Wireless Communication}.\hskip
  1em plus 0.5em minus 0.4em\relax Cambridge University Press, 2004.

\bibitem{meshkati-jsac-2006}
F.~Meshkati, M.~Chiang, H.~V. Poor, and S.~C. Schwartz, ``{A game-theoretic
  approach to energy-efficient power control in multicarrier {CDMA} systems},''
  \emph{IEEE JSAC}, vol.~24, no.~6, pp. 1115--1129, 2006.

\bibitem{EricssonWP12}
\BIBentryALTinterwordspacing
Ericsson, ``Heterogeneous networks,'' \emph{White paper}, Feb. 2012. [Online].
  Available:
  \url{http://www.ericsson.com/res/docs/whitepapers/WP-Heterogeneous-Networks.pdf}
\BIBentrySTDinterwordspacing

\bibitem{bloem-gamecomm-2007}
M.~Bloem, T.~Alpcan, and T.~Ba{\c{s}}ar, ``{A Stackelberg game for power
  control and channel allocation in cognitive radio networks},'' in
  \emph{ValueTools}, Nantes, France, 2007.

\bibitem{goodman-pcomm-2000}
D.~Goodman and N.~Mandayam, ``Power control for wireless data,'' \emph{IEEE
  Personal Communications}, vol.~7, pp. 48--54, 2000.

\bibitem{meshkati-spmag-2007}
F.~Meshkati, H.~V. Poor, and S.~C. Schwartz, ``{Energy-Efficient Resource
  Allocation in Wireless Networks},'' \emph{IEEE Signal Processing Magazine},
  vol.~24, no.~3, pp. 58--68, 2007.

\bibitem{LeeSC05}
J.~Lee, R.~V. Sonalkar, and J.~M. Cioffi, ``A multi-user power control
  algorithm for digital subscriber lines,'' \emph{IEEE Communications Letters},
  vol.~9, no.~3, pp. 193--195, 2005.

\bibitem{ElGamal2008}
L.~Lai and H.~E. Gamal, ``{The Water-Filling Game in Fading Multiple-Access
  Channels},'' \emph{IEEE Transactions on Information Theory}, vol.~54, no.~5,
  pp. 2110--2122, 2008.

\bibitem{basar-jota-2002}
T.~Ba\c{s}ar and R.~Srikant, ``{A stackelberg network game with a large number
  of followers},'' \emph{Journal of Optimization Theory and Applications}, vol.
  115, no.~3, pp. 479--90, 2002.

\bibitem{Hierarchical-Game-CR-Dusit-JSAC12}
Y.~Xiao, G.~Bi, D.~Niyato, and L.~DaSilva, ``A hierarchical game theoretic
  framework for cognitive radio networks,'' \emph{Selected Areas in
  Communications, IEEE Journal on}, vol.~30, no.~10, pp. 2053--2069, 2012.

\bibitem{Buzzi-11-Game-EE-CDMA}
S.~Buzzi and D.~Saturnino, ``A game-theoretic approach to energy-efficient
  power control and receiver design in cognitive {CDMA} wireless networks,''
  \emph{Selected Topics in Signal Processing, IEEE Journal of}, vol.~5, no.~1,
  pp. 137--150, 2011.

\bibitem{Zappone-EE-CDMA-relay-11}
A.~Zappone, S.~Buzzi, and E.~Jorswieck, ``Energy-efficient power control and
  receiver design in relay-assisted {DS/CDMA} wireless networks via game
  theory,'' \emph{IEEE Communications Letters}, vol.~15, no.~7, pp. 701--703,
  2011.

\bibitem{Bacci-13-Game-EE-OFDMA}
G.~Bacci, L.~Sanguinetti, M.~Luise, and H.~Poor, ``A game-theoretic approach
  for energy-efficient contention-based synchronization in ofdma systems,''
  \emph{IEEE Transactions on Signal Processing}, vol.~61, no.~5, pp.
  1258--1271, 2013.

\bibitem{Majed-GC-EE}
Y.~Hayel and M.~Haddad, ``A stackelberg approach for energy efficient
  multi-carrier systems,'' in \emph{Global Communications Conference
  (GLOBECOM), 2012 IEEE}, Dec 2012.

\bibitem{Majed-TVT2014}
M.~Haddad, Y.~Hayel, and O.~Habachi, ``Spectrum coordination in energy
  efficient cognitive radio networks,'' \emph{to appear in IEEE Transactions on
  Vehicular Technology}, 2014.

\bibitem{Femto08Survey}
V.~Chandrasekhar, J.~Andrews, and A.~Gatherer, ``Femtocell networks: a
  survey,'' \emph{IEEE Communications Magazine}, vol.~46, no.~9, pp. 59--67,
  2008.

\bibitem{Majed_wiopt14}
M.~Haddad, P.~Wiecek, O.~Habachi, and Y.~Hayel, ``A game theoretic analysis for
  energy efficient heterogeneous networks,'' in \emph{{WiOpt}}, Hammamet,
  Tunis, May 2014.

\bibitem{rodriguez-globecom-2003}
V.~Rodriguez, ``An analytical foundation for resource management in wireless
  communication,'' in \emph{IEEE Global Telecommunications Conference}, vol.~2,
  Dec 2003.

\bibitem{gaoning_infocom11}
G.~He, S.~Lasaulce, and Y.~Hayel, ``Stackelberg games for energy-efficient
  power control in wireless networks,'' in \emph{INFOCOM, Proceedings IEEE},
  April 2011.

\bibitem{Jondral_pooling}
T.~Weiss and F.~Jondral, ``Spectrum pooling: an innovative strategy for the
  enhancement of spectrum efficiency,'' \emph{Communications Magazine, IEEE},
  vol.~42, no.~3, pp. 8--14, Mar 2004.

\bibitem{tradeoff_green_commag11}
Y.~Chen, S.~Zhang, S.~Xu, and G.~Y. Li, ``{Fundamental tradeoffs on green
  wireless networks},'' \emph{Communications Magazine, IEEE}, vol.~49, no.~6,
  pp. 30--37, Jun. 2011.

\bibitem{MukherjeeM14}
\BIBentryALTinterwordspacing
S.~Mukherjee and S.~K. Mohammed, ``On the energy-spectral efficiency trade-off
  of the mrc receiver in massive mimo systems with transceiver power
  consumption,'' \emph{Arxiv}, 2014. [Online]. Available:
  \url{http://arxiv.org/abs/1404.3010}
\BIBentrySTDinterwordspacing

\bibitem{DengRCZZL13}
L.~Deng, Y.~Rui, P.~Cheng, J.~Zhang, Q.~T. Zhang, and M.~Li, ``A unified energy
  efficiency and spectral efficiency tradeoff metric in wireless networks.''
  \emph{IEEE Communications Letters}, vol.~17, no.~1, pp. 55--58, 2013.

\bibitem{wcnc/AminMEH12}
R.~Amin, J.~Martin, A.~Eltawil, and A.~Hussien, ``Spectral efficiency and
  energy consumption tradeoffs for reconfigurable devices in heterogeneous
  wireless systems,'' in \emph{IEEE WCNC}, April 2012.

\bibitem{JoungESTJSAC2014}
J.~Joung, C.~K. Ho, and S.~Sun, ``Spectral efficiency and energy efficiency of
  ofdm systems: Impact of power amplifiers and countermeasures,'' \emph{IEEE
  Journal on Selected Areas in Communications}, vol.~32, no.~2, pp. 208--220,
  2014.

\bibitem{TsilimantosGJ13}
\BIBentryALTinterwordspacing
D.~Tsilimantos, J.-M. Gorce, and K.~Jaffr{\`e}s-Runser, ``Spectral and energy
  efficiency trade-off with joint power-bandwidth allocation in ofdma
  networks,'' \emph{Arxiv}, 2013. [Online]. Available:
  \url{http://arxiv.org/abs/1311.7302}
\BIBentrySTDinterwordspacing

\end{thebibliography}

\appendix

\subsection{Proof of Proposition \ref{prop:CSC}}\label{app:CSC}
\begin{proof}
First note that when each of the players uses different carrier, player $n$ choosing carrier $k$ may at most obtain utility $\frac{g_n^k}{\sigma^2}$, which is achieved by allocating power $p_n^k=\frac{\gamma^*\sigma^2}{g_n^k}$ to this carrier. If at least two players allocate some power to the same carrier $k$, the utility of each of them can only decrease, thus the highest utility obtainable in the game for player $n$ (if all others cooperate to maximize his utility) is $\max_{k\leq K}\frac{g_n^k}{\sigma^2}$.

Now suppose that there exists $\alpha$ such that $\rho_{\pi(n)}(n)>\alpha$ for $n=1,\ldots,N$. Then at step $j$ of the $\pi$-CSC Algorithm user $\pi(j)$ chooses one of the carriers that are not chosen yet. As there are only $j-1$ carriers already chosen, he will choose one of $j$ carriers, whose $g_{\pi(j)}^k$ are the greatest. For each of them
$$g_{\pi(j)}^k\geq\rho_{\pi(j)}(j)\max_{k\leq K}g_{\pi(j)}^k>\alpha\max_{k\leq K}g_{\pi(j)}^k.$$
The utility obtained by $\pi(j)$ when he will allocate power $\overline{p}_{\pi(j)}^k=\frac{\gamma^*\sigma^2}{g_{\pi(j)}^k}$ to such a $k$ will thus be
$$\frac{g_{\pi(j)}^k}{\sigma^2}>\alpha\max_{k\leq K}\frac{g_{\pi(j)}^k}{\sigma^2},$$
which is $\alpha$ times the biggest utility obtainable for player $\pi(j)$ in the game.\\
\end{proof}

\subsection{Proof of Proposition \ref{prop:OCSC}}\label{app:OCSC}
\begin{proof}
First note that if Algorithm \ref{alg:alpha_pi_choice} does not give an output $(\alpha^*,\pi^*)$ with $\pi^*\neq\boldsymbol{0}$, we can take $\alpha^*=0$ and $\pi^*=(1,2,\ldots,N)$. This is because for $\alpha^*=0$ every $\rho_n(l)$ will be bigger than $\alpha^*$. Thus any ordering of the players $\pi^*$ will satisfy $\rho_{\pi^*(l)}(l)\geq\alpha^*$ for every $l=1,\ldots,N$ and by Proposition \ref{prop:CSC} and Corollaries \ref{CSC:equilibrium} and \ref{CSC:soc_opt} also statements 1), 2) and 3) of the proposition.

Next, see also that Algorithm \ref{alg:alpha_pi_choice} is always terminated at some step, as after each passage through points 1)--3) of the algorithm, $\overline{\alpha}-\underline{\alpha}$ decreases twice, thus at most after $\log_2(\delta^{-1})$ passages, $\overline{\alpha}-\underline{\alpha}<\delta$, which stops the algorithm.

Now suppose that $\alpha^*$ and $\pi^*$ are nonzero outputs of Algorithm \ref{alg:alpha_pi_choice}. Then for every $l=1,\ldots,N$, $\rho_{\pi^*(l)}(l)\geq\alpha^*$, as by construction of $\pi^*$, for each $l^*$ chosen in the algorithm, $\alpha^*\leq\rho_{\pi^*(l)}(l^*)\leq\rho_{\pi^*(l)}(l)$, where the last inequality is due to the facts that $l\leq l^*$ and that $\rho_n(l)$ are decreasing in $l$. Again, we can apply Proposition \ref{prop:CSC} and Corollaries \ref{CSC:equilibrium} and \ref{CSC:soc_opt} to obtain statements 1), 2) and 3) of the proposition also in this case.

Now suppose that the first part of the $\delta$-OCSC Algorithm (Algorithm \ref{alg:alpha_pi_choice}) was terminated by condition $(\alpha^*=)\underline{\alpha}>\frac{1}{1+\gamma^*}$, and let $(\overline{p}_1,\overline{p}_2,\ldots,\overline{p}_N)$ be the power control vector chosen in the second part of the algorithm. We will show that these controls form an equilibrium in the game. Suppose they do not, that is -- there exists a player $n$ who can gain by deviating form it. First note that he will not deviate to any of the carriers that are not used by anyone, as $\pi^*$-CSC algorithm assures that player $n$ always chooses his best available carrier at the time decision is made. Obviously the set of unoccupied carriers may only shrink afterwards, so any carrier that is idle upon the termination of the algorithm can only decrease the utility for player $n$. Thus suppose that player $n$ changes his carrier to some carrier $k^*$, used by some other player $j$ using power $p_j^{k^*}=\frac{\gamma^*\sigma^2}{g_j^{k^*}}$. By Proposition \ref{prop:follower-power} (adapted to the $N$-player case) the biggest utility for player $n$ that can be obtained on that carrier will be for power
\begin{eqnarray}
p_n^{k^*}&=&\frac{\gamma^*(\sigma^2+g_j^{k^*}p_j^{k^*})}{g_n^{k^*}}
=\frac{\gamma^*(\sigma^2+g_j^{k^*}\frac{\gamma^*\sigma^2}{g_j^{k^*}})}{g_n^{k^*}}\nonumber\\
&=&\frac{\gamma^*\sigma^2(1+\gamma^*)}{g_n^{k^*}}.
\end{eqnarray}
This utility will be thus equal to
$$\frac{\gamma^*}{p_n^{k^*}}=\frac{g_n^{k^*}}{\gamma^*\sigma^2(1+\gamma^*)}.$$
On the other hand the utility obtained by $i$ on carrier $\kappa(n)$ he currently uses is $\frac{g_n^{\kappa(n)}}{\gamma^*\sigma^2}$ and by already proved part 1) of Proposition \ref{prop:OCSC} not worse than
$$\alpha^*\max_{k\leq K}\frac{g_n^k}{\gamma^*\sigma^2}>\max_{k\leq K}\frac{g_n^k}{\gamma^*(1+\gamma^*)\sigma^2}\geq\frac{g_n^{k^*}}{\gamma^*\sigma^2(1+\gamma^*)},$$
which ends the proof of part 4) of the proposition.

Finally let $\tilde{g}_n^k$ be defined by
$$\tilde{g}_n^k=\rho_n(k)\max_{l\leq K}g_n^l.$$
Obviously $\tilde{g}_n^k$ give the same values of $\rho_n(k)$ as $g_n^k$. Note also that, for each of the players, the biggest value of $\tilde{g}_n^k$ is for $k=1$, the second biggest for $k=2$ and so on. Thus, for such $\tilde{g}_n^k$,
\begin{equation}
\label{rho_star}
\rho_n^k=\rho_n(k).
\end{equation}
Obviously, for such values of $\tilde{g}_n^k$, each of the players will always be better off choosing one of the carriers $1,\ldots,N$ than any other ones. Thus we can assume that in our problem $N=K$. Now suppose that, for some $\delta>0$, there exists a power control in the game, inducing spectrum coordination and such that each of the players receives more than $(\alpha^*+\delta)\max_{k\leq K}\frac{g_n^k}{\gamma^*\sigma^2}$, which is $\alpha^*+\delta$ times the biggest possible value of his utility (where $\alpha^*$ is obtained in Algorithm \ref{alg:alpha_pi_choice} for $\delta$). Let $k^*(n)$ be the carrier chosen by player $n$ in that control. Since by 1) of Proposition \ref{prop:OCSC} it could not be found by the $\delta$-OCSC Algorithm, no ordering of the players $\pi$ can be found such that
\begin{equation}
\label{viol_eq}
\rho_{\pi(n)}(n)\geq\alpha^*+\delta
\end{equation}
for every $n=1,\ldots,N$. Let us take $\pi$ such that $k^*(\pi(n))=n$ for $n=1,\ldots,N$ (since $K=N$ and we have spectrum coordination, this is well defined). Let $i^*$ denote $i$ for which (\ref{viol_eq}) is not satisfied. For such an $i^*$ we have
$$\frac{g_{\pi(i^*)}^{k^*(\pi(i^*))}}{\max_{k\leq K}g_{\pi(i^*)}^k}=\rho_{\pi(i^*)}^{k^*(\pi(i^*))},$$
which by (\ref{rho_star}) equals
$$\rho_{\pi(i^*)}(k^*(\pi(i^*)))=\rho_{\pi(i^*)}(i^*)<\alpha^*+\delta$$
and thus there exists a player $j=\pi(i^*)$ such that
$$g_j^{k^*(j)}<(\alpha^*+\delta)\max_{k\leq K}g_j^k$$
which implies that his utility is
$$\frac{g_j^{k^*(j)}}{\gamma^*\sigma^2}<(\alpha^*+\delta)\max_{k\leq K}\frac{g_j^k}{\gamma^*\sigma^2},$$
contradicting the assumption that each of the players receives more than $(\alpha^*+\delta)\max_{k\leq K}\frac{g_n^k}{\gamma^*\sigma^2}$.
\end{proof}

\end{document}